\documentstyle [12pt]{article}
\begin{document}
\title{Unconditionally Secure Quantum Bit Commitment is Simply Possible} 
\author{Arindam Mitra
\\Lakurdhi, Tikarhat Road, Burdwan, 713102.\\
 West Bengal, India.}

\maketitle

\begin{abstract}\bf
Mayers, Lo and Chau proved unconditionally secure quantum bit commitment 
is impossible. It is shown that their proof is valid only for
a particular model of 
quantum bit commitment encoding, in general it does not hold good.
A different model of unconditionally secure
quantum bit commitment -
both entanglement
and disentanglement-based - is presented.
Even cheating can be legally proved with some legal evidences.
Unconditionally secure quantum bit commitment
can be established on the top of unconditionally secure
quantum coin tossing, which is
also claimed to be two-way impossible.

\end{abstract}

\newpage 

\noindent 

The task of quantum key distribution (QKD)  
is to provide  identical sequence of random bits  for 
two distant parties - sender and receiver. Its security
against eavesdropping is guaranteed by quantum mechanics.
But the question is:
Will they always get identical sequence ?  From
conventional  quantum key distribution (QKD) protocols [1,2], they
cannot have identical sequence if
one of them  becomes dishonest. 
It seems to be a non-issue.
If they want to communicate
secretly there is no reason of being dishonest.
One may even argue that secure communication between mistrusted
parties is itself meaningless and therefore, honesty is the best policy
in secure communication. 
But, in conventional QKD protocols, dishonesty is allowed by the protocol
itself. This is a new thing. \\

To elucidate the issue, let us recall the BB-84 QKD protocol [1].
Like all other conventional QKD protocols it also works
on two-step process. In the first step,
sender transmits a sequence of $0^{\circ}$,
$90^{\circ}$, $45^{\circ}$and $135^{\circ}$ polarized single photons.
The $0^{\circ}$ and $45^{\circ}$ single photons represent bit 0 and
$90^{\circ}$and $135^{\circ}$ single photons represent bit 1.
Receiver could recover the bit values if  sender gives the required
information (basis of measurements) regarding the bit values.
In the second step, sender reveals the required
information  to  receiver. The problem is,
sender can flip the bit value by changing
the required information although he committed the bit value in the first step.
This is  cheating.\\

This particular type of cheating can be described as $180^{\circ}$ shift
from commitment. This shift may be accepted if receiver does not get
ultimately cheated.
Bennett and Brassard were aware about the problem and they
observed that their BB-84 protocol is totally insecure against
cheating if sender uses suitable entangled states instead of the said BB-84 states.
To overcome this difficulty, the idea of bit commitment surfaced
in the early 90's. It was anticipated that if quantum bit commitment
(QBC)
is established on the top of QKD scheme then cheating could be detected.
As if, in cryptographic communication
quantum mechanics could resist a committed partner
to be an imposter. If secure QBC protocol is found, it was thought that
it could be the basis of other important cryptographic
schemes such as secure quantum coin tossing, secure 
quantum oblivious transfer, secure two-party quantum computation.
So, the security issue of quantum bit commitment has immense importance.\\

In 1995, on the basis of conventional model quantum cryptography,
  a  QBC protocol [3], known as BCJL scheme, was proposed and claimed to be
provably secure against all types of cheating. Mayers [4] followed by
Lo and Chau
proved [5] it incorrect. But  message of their work is that there
cannot have any  secure bit commitment protocol, although they
worked on a particular model of quantum bit commitment encoding.
Recently Kent [6] has invaded this belief. He showed that secure
classical bit commitment  protocol exists.  As the security of his protocol is based on
special theory of relativity, it is still widely believed
that their  proof is  valid for all unknown
quantum
bit commitment protocols [7],
which will not use relativity to ensure security against
cheating. If it be so, in cryptography
  relativity wins over quantum mechanics. 
We shall see, that the belief - quantum cryptography is  too weak to
realize bit commitment encoding- is misplaced.\\

We shall first discuss why their proof cannot be considered as a generalized result.
Recall the reasoning of complete
cheating. Complete cheating is
possible when two density matrices associated with bit 0 and 1  are
same i.e $\rho_{0}=\rho_{1}$. Because of this equivalence of
two density matrices, using entanglement,
sender, after transmitting  the state $\vert 0 \rangle$,
corresponding to bit 0, 
 can alone apply  unitary transformation
U to convert $ \vert 0 \rangle $ to $\vert 1 \rangle$, corresponding to bit 1
and vice versa,
keeping the receiver in dark about this transformation. But it does not necessarily mean
whenever  $\rho_{0}=\rho_{1}$ cheating will be possible and successful. \\

Consider the following simple  quantum coding technique [8-9]. \\

\noindent Bit $ 0 \longrightarrow \left\{\psi, \, \, \phi, \, \,
\psi, \, \, \phi, \, \,
\psi, \, \psi, \,
\phi, \, \,
\phi, \,\, \psi,
\,\, \phi,.......................\right\}$\\
Bit $ 1 \longrightarrow \left\{\phi, \, \, 
\phi, \, \,
\psi, \, \,
\psi, \, \, \phi, \, \,
\psi, \, \, \psi, \, \,
\phi, \, \, \phi, \, \, 
\psi,......................\right\}$ .\\
 
These are two reasonably large sequences
of two nonorthogonal quantum states $\psi$ and $\phi$ (
they are strictly not orthogonal because it will be classical encoding
with quantum states ).
Suppose these are two sequences  of $0^{\circ}$ and $45^{\circ}$
(1:1) polarized  single photons.
So $\rho_{0} = \rho_{1}$. Information regarding the
above two sequences is shared between sender and
receiver. Here cheating is not possible as receiver can 
alone recover the bit value from the information they initially shared. 
The simple method of recovery of bit values can be like this: Bob uses 
analyzer at $0^{\circ}$ and $45^{\circ}$ orientations for his measurements
and wants to recover the bit values from exactly one half of the
transmitted   sequences without missing to detect a single state.
In the first 50\% events Bob
 measures  according to the first (given above)
sequence and in the last 50\% events he measures according to
the second sequence properly using his analyzers.
Therefore, always he could {\em statistically} and {\em deterministically}
 recover the exact half
of any of the above two sequences. If exact 
first half of the first sequence is recovered then bit is {\em deterministically}
 0. Similarly if exact last half of the second sequence is recovered then the bit
is {\em deterministically} 1.
But this is a cheating-free single step
QKD protocol not the two-step quantum bit commitment protocol.  Similarly
  our single step
 entanglement- based QKD protocol [9] can be considered as
 a cheating-free protocol.
   It perhaps implies that
 cheating was possible in conventional QBC as because they
 did not share information of the two density matrices
 not because of the equivalence of
 density matrices though their encoding does not allow to do so.
  So sharing of information
 can be a precondition to have a secure QBC. But
 this precondition is not enough to realize two-step bit commitment
 encoding. Next  we shall
 see, single step cheating-free QKD protocol can be simply modified into
two-step protocol to realize unconditionally secure 
quantum bit commitment. First we shall present a two-step
entanglement-based QBC protocol.\\

Suppose Alice has n pairs of EPR particles. Taking one particle of each pair,
she   arranges them in a particular fashion and taking the partner particles
she arranges them  in another way with the help of quantum memory. Suppose  the two arrangements are :\\

\noindent $S_{0} 
= \left\{A,\,\, B,\,\,C,\,\,D,\,\,E,\,\,F,\,\,G,\,\,H,
.........\right\}$, \\
$s_{0} = \left\{b,\,\,f,\,\, g,\,\,a,\,\,e,\,\,h,\,\,d,\,\,c,.............
\right\}$\\

Here   capital and its  small letter stand for an entangled pair.
That is, particle "A" and  "a" form an EPR pair and 
particle "B" and "b" form another EPR pair and so on. 
 These two arrangements represent bit 0.
To represent bit 1, similarly she can arrange them in another two 
different ways:\\

\noindent
$S_{1} 
= \left\{M,\,\, N,\,\,O,\,\,P,\,\,Q,\,\,R,\,\,S,\,\,T,
.........\right\}$,\\ 
$s_{1} = \left\{s,\,\,o,\,\,n,\,\,p,\,\,t,\,\,q,\,\,m,\,\,r,
..........\right\}$.\\

To avoid confusion we have used two sets of capital and
small letters to denote entangled pairs.
The entangled state can be represented as,\\
$\vert\psi\rangle_{i,j} =
1/\sqrt 2 (\vert\uparrow\rangle_{i}\vert\downarrow\rangle_{j} -
\vert\downarrow\rangle_{i}\vert\uparrow\rangle_{j})$\\
where i and j denote the position of the EPR particles in $S_{0/1}$ and
$s_{0/1}$ respectively.
The above information about the two arrangements 
is secretly 
shared between them.\\

Bit commitment encoding can be executed in  two-step process.
 In    the first step, Alice commits bit 0 by sending  $S_{0}$ and in the second step,
she reveals the required information just by sending $s_{0}$ . Similarly she can commit
bit 1 by sending $S_{1}$ in the 1st step and reveals its value by sending
 $s_{1}$ in the second step.
 Instead of directly sending the 2nd sequence, the
results of measurements on 2nd sequence in a pre-committed basis can also  be revealed.
From the first incoming sequence $S_{0}$ or $S_{1}$,
Bob cannot recover the bit
values. But he can alone recover  Alice's committed bit when he will get the
partner sequence $s_{0}$ or $s_{1}$. He can measure the spin
in a fixed direction. Measurements on the two sequences   
of EPR particles will
produce  correlated data. Bob's task is to recover  the
bit values   from these data and initially shared data.
If dishonest Alice sends $s_{1}$ after
$S_{0}$ or $s_{0}$ after $S_{1}$,  then Bob could not identify
any of the bit values because EPR correlation will be lost in the case of
cheating.  Thus cheating will be exposed.  \\

The protocol, described above, is an entanglement-based QBC protocol.
Using two  sequences of deliberately prepared 
superposition states and following the same operational procedure 
disentanglement based  QBC protocol can be given.
Suppose the superposition states are:\\  

\noindent$\vert A \rangle_{i} 
= 1/\sqrt 2 (\vert\leftrightarrow\rangle^{\bf r}_{0} + \vert\leftrightarrow\rangle^{\bf s}_{0})$\\
$\vert B \rangle_{i} = 1/\sqrt 2 (\vert\leftrightarrow\rangle^{\bf r}_{0} 
+ \vert\updownarrow\rangle^{\bf s}_{0})$. \\
$\vert C \rangle_{i} = 1/\sqrt 2 (\vert\leftrightarrow\rangle^{\bf r}_{1} +
\vert{\nearrow\!\!\!\!\!\!\swarrow}\rangle^{\bf s}_{1}$\\
$\vert D \rangle_{i} = 1/\sqrt 2 (\vert\leftrightarrow\rangle^{\bf r}_{1} + 
\vert{\nwarrow\!\!\!\!\!\!\searrow}\rangle^{\bf s}_{1}$.\\

The sequence ($Q_{0}$) of the states $\vert A \rangle_{i}$ and
$\vert B \rangle_{i}$
represent
bit value 0 and the sequence ($Q_{1}$) of states
$\vert C \rangle_{i}$ and $\vert D
\rangle_{i}$ represents bit value 1.
The preparation procedure of these
superposition states has been discussed in  ref 8.
To commit
the bit value, say 0, Alice in the 1st step,   splitting each
state $\vert\,\,\, \rangle$ of the shared sequence ($Q_{0}$) of states
sends the  sequence ($S_{0}$) of the truncated state $\vert \,\,\,\rangle_{\bf r}$
which does not contain the bit value.
The path {\bf s } is the bit-carrying path.
Alice keeps the remaining
 sequence ($s_{0}$) of states $\vert \,\,\,\rangle_{\bf s}$ in quantum memory (using delay).
In the second step, Alice transmits the sequence ($s_{0}$) of the
states $\vert \,\,\,\rangle_{\bf s}$ to Bob. Note that,  positions
of the complete state and truncated states in their respective sequences are same.
In that sense
 $Q_{0} = S_{0} = s_{0}$ and $Q_{1} = S_{1} = s_{1}$, where $S_{1}$ and
 $s_{1}$ are wave fuction-splitted sequences representing the bit 1.
Bob can alone recover the bit values from the second sequence
because it carries the bit values.
The simple method of recovery of
bit value from the second sequence is discussed below. \\

Suppose in the sequence $Q_{0}$  the
 $ \vert A \rangle_{i}$s are at odd positions and $\vert B \rangle_{i}$s are at 
even positions but   $ \vert C \rangle_{i}$s and  $ \vert D \rangle_{i}$s
 have no such regularity in $Q_{1}$. Now  Bob uses
$90^{\circ}$ analyzer to measure on the second sequence  of states
$\vert \,\,\rangle_{s}$. 
He gets a sequence of "yes" and "no" results. If the results 
"yes" come only at even positions, then the bit is 0.
If Alice transmits
$s_{1}$ after transmitting $S_{0}$ or $s_{0}$ after $S_{1}$
Bob will certainly be aware of such improper execution
of the protocol.
Bob will have  to go through the dual measurements on
both the sequences(need not be at the same time),
if he wants to know whether Alice is cheating
or not.
The probability of dual occurrence of result "yes"  is given in table 1,
considering Alice transmits $s_{0}$ after $S_{1}$ or $s_{1}$ after $S_{0}$
and Bob uses both analyzers at $0^{\circ}$.  One cannot get
double "yes" from a single particle. It implies that not only Bob but also
any third party, who does not know anything about their shared information,
could spot the cheating. It implies Bob could prove before the court that
Alice tried to cheat him provided some legal evidences help him.
This is also true for entanglement-based QBC,
but Bob has to reveal their shared secret before the court.
Of course before going to the court Bob has to be certain that there
is no meaningful correlation in the data sets since Alice can transmit
her two private sequences at random to defame Bob before the court by
disapproving the Bob's revealed data as their shared data.
This is an interesting thing - we are tempted to say that honesty is the only policy
in quantum communication.
 \\

The above protocol
reveals another interesting thing: due to superposition principle
it is possible to commit the bit value without sending the actual
bit-carrying
part of the wave function. The situation can be thought as a case of
{\em commitment prior
to commitment}. On the other hand,
in  our entanglement-based
QBC protocol both first and second sequences are required to recover the bit value.
So the significant difference between our entanglement-based and
disentanglement-based QBC is that  cheating is possible
(although it will be unsuccessful) in
disentanglement-based QBC but cheating is totally impossible
in entanglement-based
QBC.  \\

In the above two schemes, bit commitment encoding is two-step process.
The QBC  can be realized through  multi-step procedure.  Alice can commit
through many steps and reveals the commitment after that (it can also be
thought as a single-step commitment followed by multi-step disclosure ). 
Yet the commitment is secure. The encoding is same except we need 
higher dimensional Hilbert space (for fixed n ) to execute
multi-step QBC. As for example, they can take GHZ state 
$\vert\psi\rangle_{GHZ} =1/\sqrt 2 (\vert \uparrow_{G} \,
 \uparrow_{H}\, \uparrow_{Z}\rangle  +\vert \downarrow
_{G}\,  \downarrow _{H} \, \downarrow_{Z}\rangle$)[11].
The n copies of three entangled particles ( denoted by G, H, and Z)
can be arranged in three different ways
to represent bit 0. The arrangements are denoted by 
$G_{0}$, $H_{0}$ and $Z_{0}$.
Similarly Alice can arrange them in another three different ways,
denoted by $G_{1}$, $H_{1}$ and $Z_{1}$, to represent
bit 1.
Alice in the first step commits bit 0 by sending $G_{0}$ and reveals
the commitment by sending $H_{0}$ and $Z_{0}$ in  the next two steps. Similarly
she can commit the bit 1.
If they want to have a multi-step disentanglement-based
QBC scheme they can use a linear chain of  superposition state 
of our earlier protocol.
 For  three-step disentanglement-based QBC,  the superposition states
 are (see ref 8): \\

\noindent $\vert A \rangle_{i} = 
1/\sqrt 3 (\vert\leftrightarrow\rangle^{\bf r}_{0} + 
\vert\leftrightarrow\rangle^{\bf t}_{0} + \vert\leftrightarrow\rangle^{\bf s}_{0} )$\\
$\vert B \rangle_{i} = 1/\sqrt 3 (\vert\leftrightarrow\rangle^{\bf r}_{0} + 
\vert\leftrightarrow\rangle^{\bf t}_{0} + \vert\updownarrow\rangle^{\bf s}_{0} ) $\\
\noindent $\vert C \rangle_{i} 
= 1/\sqrt 3 ( \vert\leftrightarrow\rangle^{\bf r}_{1} + 
\vert\leftrightarrow\rangle^{\bf t}_{1} + 
\vert{\nearrow\!\!\!\!\!\!\swarrow}\rangle^{\bf s}_{1}$ \\
$\vert D \rangle_{i} = 1/\sqrt 3 ( \vert\leftrightarrow\rangle^{\bf r}_{1}
+ \vert\leftrightarrow\rangle^{\bf t}_{1} + 
\vert{\nwarrow\!\!\!\!\!\!\searrow}\rangle^{\bf s}_{1}$ .\\

Again sequence ($Q_{0}$) of states $\vert A \rangle_{i}$ and $\vert B \rangle_{i}$ 
represent bit 0 and sequence ($Q_{1}$) of states 
$\vert C \rangle_{i}$ and $\vert D \rangle_{i}$ represent bit 1.
Alice commits the bit value by sending
the sequence  of states $\vert \leftrightarrow\rangle^{\bf r}$ and reveals the
commitment by
sending the sequence of states $\vert \,\,\,\rangle^{\bf t}$ first 
and then by sending the actual bit-value-carrying sequence of states 
$\vert \,\,\,\rangle_{\bf s}$.\\ 

To prove unconditional security the effect of noise is
excluded.
We shall consider it.
Due to noise some of the Bob's measured data will be
corrupted. 
Manipulating noise (bringing noise level down)
Alice can execute the protocol dishonestly up to the noise level.
Nevertheless Bob can statistically faithfully 
recover the bit value in presence of  noise.
The main advantage of initial sharing of information of bit preparation is that
we will not have to be worried about any unknown attacks. 
Note that, sharing means pre-commitment and this can give security against
cheating
even for unknown attacks. The BCJL scheme [3] failed because presently
known attack was not clearly known to
the authors.\\

In our alternative QBC protocols, the probability of the success in cheating
is always zero. Security does not depend on time, space, technology,
noise, and unknown attacks.  Therefore,  protocols can be safely claimed 
as absolutely 
secure protocols against cheating. It can be mentioned that security is not
coming from quantum mechanics; it only allows us to perform quantum
bit commitment. It is interesting to note, conventional QBC protocol totally
fails because of entanglement. The same entanglement provides us secure QBC,
although  bit commitment is not the problem of alternative QKD.
Regarding bit commitment issue, entanglement is not our enemy rather our
friend indeed.\\

Coin tossing is another important cryptographic primitive:
two distant mistrusted parties want to generate
faithful random bits to authenticate the channel
either by classical coin tossing or by quantum coin tossing (QCT).
We can think of two types of coin tossing  -ideal
and nonideal. It should be mentioned that  if one of them
does not want to simulate the real coin tossing there is no physical law
which can compel him/her to do so.
The question is, how far the can the generated bits  be considered  secure
against cheating ?
Very simple unconditionally secure classical coin tossing protocol exists [13]
Lo and Chau  claimed that  [12]   secure
ideal QCT is impossible.
Their proof is based on the assumptions:
ii) shared entanglement cannot be proven
genuine ii) entanglement is a necessary condition
for secure ideal QCT.
We have shown  how to check [9]
the authenticity of the shared entangled states. Therefore,
simulation of
ideal QCT
is simply possible.
It is well known that  QCT can be based on
QBC protocol. So, we are getting second QCT from our QBC.
 In addition to that, our QKD protocols
are  basically  QCT protocols.
Alternative QCT protocols
can be ideal or  non-ideal QCT protocols.
That is, every bits are secure.
These three types of QCT
are unconditionally secure against cheating . Yet they cannot be used for
authentication until they are proved absolutely secure in presence of noise.
\\

The power of different cryptographic primitives is itself a subject of interest.
Recently Kent has claimed that QCT cannot be built on the top of
QBC and therefore it is weaker than QBC. We have
already seen that our QKD can be thought as QCT on which we can implement
our QBC.
So Kent's proof cannot encompass our model.
We have seen that all QKD/QCT protocols
are not QBC protocol. Is the reverse true ? The reverse will not be true
if there is regularity in each of the two operating sequences.
This type of QBC
cannot be used as secure  QCT for authentication. We conclude:
1. unconditionally secure QBC can be implement on the top of
unconditionally secure QCT.
2. unconditionally secure QCT can be implement on the top of
unconditionally secure QBC.
3. Every QBC is not QCT scheme.
4. Every QCT is not QBC scheme.\\

Yao has proved [14] that secure quantum oblivious is possible if
secure quantum bit commitment is found. On the other hand Killian [15]
has proved that secure oblivious scheme can be the basis of
secure one-sided two-party computation.
Applying classical reduction theory it has been argued that secure quantum
computation scheme can be derived from the secure quantum
bit commitment scheme. Now we have got secure quantum bit commitment
scheme, can we hope for such secure quantum computation scheme ?
The problem is, Lo has already proved [16] that secure one-sided two-party
quantum computation is impossible. We are in a fix. Either Lo's proof
is not a generalized result or the chain of logic is partly or totally incorrect. At least
both cannot be right. This puzzle deserves further investigation. \\

There is another misleading analysis on  conventional
quantum cryptographic model. For a particular eavesdropping attack,
it is stated   that optimal information
gain of the eavesdropper versus introduced error by him/her is bounded by the laws of
quantum mechanics.
This is true if there is only one eavesdropper. If we consider
many eavesdroppers then optimal gain of information of
any eavesdropper will depend on
the co-operation of other eavesdroppers
which quantum mechanics cannot dictate. Considering
the many eavesdropping issue
one can even lead to the conclusion: two eavesdroppers are
more acceptable than one eavesdropper if one has to accept
eavesdropping and can tolerate error. But this discussion is beyond the scope of this paper.
One may wonder: why so much shortcomings ?
Perhaps topics demands so.\\

In conclusion, as QBC issue tells us how to   distribute
cheating-free  information at different time, 
it might have different  applications in public and private life.
And this  is possible
to implement within the present technology because,
 without storing the quantum
states the results of measurements can be stored and revealed
later instead of storing quantum states and  sending them later
to execute QBC.\\

I thank C. H. Bennett for  one of
his comment  on alternative QKD protocol that activated this work.\\ 

\noindent
e-mail:mitra1in@yahoo.com

\newpage

\begin{table}\bf Table 1. Joint probabilities when $DA$
at $(0^{\circ}
 :   0^{\circ})$\begin{center}  \begin{tabular}{|c|cccc|}\hline
  commited state $ \longrightarrow $ revealed state               &
$P_{(\surd^{\bf r}:\surd^{\bf s})}$ &
$P_{(\surd_{r}:\times_{\bf s})}$
& $P_{(\times^{\bf r}:\surd^{\bf s})}$
& $P_{(\times^{\bf r}:\times^{\bf s})}$        \\        \hline
$\vert{\leftrightarrow}\rangle^{\bf r}_{0} \longrightarrow                  
\vert{\nearrow\!\!\!\!\!\!\swarrow}\rangle^{\bf s}_{1}$
 & $1/32$ & $1/32$ &  $1/32$  &  $1/32$  \\
$\vert{\leftrightarrow}\rangle^{r}_{0} \longrightarrow                  
\vert{\nwarrow\!\!\!\!\!\!\searrow}\rangle^{\bf s}_{1}$
 & $1/32$ & $1/32$  &  $1/32$  &  $1/32$    \\

$\vert{\leftrightarrow}\rangle^{\bf r}_{1}  \longrightarrow                 
\vert{\leftrightarrow}\rangle^{s}_{0}$ & $1/16$ &  $1/16$  &$1/16$ &  $1/16$    \\  
$\vert{\leftrightarrow}\rangle^{\bf r}_{1}  \longrightarrow
\vert{\updownarrow}\rangle^{\bf s}_{0}$ & 0 & $1/16$  & 0 & $1/16$  \\ \hline \end{tabular}\end{center}
 \small\end{table}

\end{document}